\documentclass[aps,prd,preprint,showkeys,showpacs,unsortedaddress,superscriptaddress]{revtex4-1}
\usepackage{amsmath, amsfonts, amssymb, mathrsfs}
\usepackage{amsthm}
\usepackage{graphicx,color}
\usepackage{bm}
\newcommand{\nn}{\nonumber}

\usepackage{setspace}
\begin{document}

\title{The post-Newtonian metric of a self-gravitating and collapsing thin spherical shell}%
\author{Wenbin Lin}\email{lwb@usc.edu.cn}
\affiliation{School of Mathematics and Physics, University of South China, Hengyang, 421001, China}
\affiliation{School of Physical Science and Technology, Southwest Jiaotong University, Chengdu, 610031, China}

\date{\today}

\begin{abstract}
We calculate the metric for a self-gravitating and collapsing infinitely-thin spherical shell under the theory of post-Newtonian approximation, and derive the post-Newtonian equation-of-motions the test particles inside and outside the shell. It is demonstrated that the dynamic spherically-symmetric system can produce the time-dependent gravitational field. 


~~\\
{Keywords: the post-Newtonian approximation, the second gravitational potential, gravitomagnetic field.}

\pacs{04.20.-q, 04.20.Cv, 04.25.Nx, 95.30.Sf}

\end{abstract}

\maketitle

\section{Introduction}

The thought experiments, for example, Einstein's elevator experiment, play important roles in the development of theoretical physics. In this work, we consider the scenario of a self-gravitating and collapsing thin spherical shell, to demonstrate that the motion of the gravitational source will produce the relativistic effects in general, especially, the radial motion of the spherically-symmetric system can induce the second gravitational potential and the gravitomagnetic field, both of which are time-dependent. 

In order to get the analytical solution for this experiment, we adopt the theory of the post-Newtonian (PN) approximation, which is the classical method to solve Einstein field equations in the weak-field and slow-motion limits~\cite{Weinberg1972,Will2011,PoissonWill2014}, to calculate the external and internal metric of the collapsing shell. 

The paper is organized as follows. Section~\ref{sec2} reviews the theory of post-Newtonian approximation. In Section~\ref{shell} we derive the 1PN external and internal metrics of the self-gravitating and collapsing thin spherical shell. In Section~\ref{dynamic equation} we give the test particles' dynamic equations outside and inside the shell respectively. Conclusion is given in Section~\ref{con}.

\section{Theory of the post-Newtonian approximation}\label{sec2}

Einstein field equations can be written as~\cite{Weinberg1972}
\begin{eqnarray}
\mathcal{R}_{\mu\nu} = - 8\pi \big(T_{\mu\nu}-\frac{1}{2}g_{\mu\nu}T^{\lambda}_{\lambda}\big)~,\label{EE}
\end{eqnarray}
where $\mathcal{R}_{\mu\nu}$ is Ricci tensor, $g_{\mu\nu}$ is the metric tensor and $T_{\mu\nu}$ is the matter energy-momentum tensor. We use the natural units in which $G\!=\!c\!=1$. Greek indices run from $0$ to $3$. Einstein summation notation is used. 

In the limits of weak-field and slow motion, Einstein field equations can be solved by the method of the post-Newtonian approximation, which has been well-established and applied to solve numerous gravitational problems~\cite{Will2011}. For the integrity of this work, here we review the basic contents of the theory of the post-Newtonian approximation, and adopt the same notations as in Weinberg's textbook~\cite{Weinberg1972}.

Let $\bar{M}$, $\bar{v}$, and $\bar{r}$ represent the typical values of mass, velocity, and distance in a non-relativistic system. In the post-Newtonian approximation, the metric is expanded in the powers of ${\bar{v}}^2$, which is roughly of the same order of the typical Newtonian potential $\bar{\phi}=\!-\bar{M}/\bar{r}$, as follows
\begin{eqnarray}\label{gmunu}
 g_{00} &=& - 1 + \stackrel{2}g_{00} + \stackrel{4}g_{00} + \cdots~, \label{g00}\\
 g_{0i} &=& \stackrel{3}g_{0i}  + \cdots~, \label{g0i}\\ 
 g_{ij} &=& \delta_{ij} + \stackrel{2}g_{ij} + \cdots~, \label{gij} 
\end{eqnarray}
where Latin indices run from $1$ to $3$. $\delta_{ij}$ is Kronecker's delta. $\stackrel{N}g_{\mu\nu}$ denotes the term in $g_{\mu\nu}$ of order ${\bar{v}}^N$. 
The corresponding energy-momentum tensor is expanded as
\begin{eqnarray}\label{Tmunu}
 T^{00} &=& \stackrel{0~~}{T^{00}} + \stackrel{2~~}{T^{00}} + \cdots~,  \label{T00PN}\\
 T^{0i} &=& \stackrel{1~~}{T^{0i}} + \cdots~, \label{T0iPN}\\ 
 T^{ij} &=& \stackrel{2~~}{T^{ij}} + \cdots~, \label{TijPN} 
\end{eqnarray}
where $\stackrel{N~~}{T^{\mu\nu}}$ denotes the term in $T^{\mu\nu}$ of order $(\bar{M}/{\bar{r}^3}){\bar{v}}^N$.\\
\indent Substituting Eqs.~(\ref{g00}) - (\ref{TijPN}) into Eq.~(\ref{EE}), and making use of the harmonic-coordinate conditions, we have
\begin{eqnarray}
&& \hskip -0.5cm \stackrel{2}{\mathcal{R}}_{00}=\frac{1}{2}\nabla^{2}\overset{2}g_{00} = -4\pi \overset{0}T{^{00}}~,\label{R002}\\
&& \hskip -0.5cm \stackrel{2}{\mathcal{R}}_{ij}=\frac{1}{2}\nabla^{2}\overset{2}g_{ij} = -4\pi \delta_{ij} \overset{0}T{^{00}}~,\label{Rij2}\\
&& \hskip -0.5cm \stackrel{3}{\mathcal{R}}_{0i}=\frac{1}{2}\nabla^{2}\overset{3}g_{0i} = 8\pi \overset{1}T{^{0i}}~,\label{R0i3}\\
&& \hskip -0.5cm \overset{4}{\mathcal{R}}_{00}=
\frac{1}{2}\big[\nabla^{2}\overset{4}g_{00}\!-\!\partial_t^2 \overset{2}g_{00} \!-\! \overset{2}g_{ij}\partial_{i} \partial_{j} \overset{2}g_{00}\!+\!(\partial_i \overset{2}g_{00})(\partial_i \overset{2}g_{00})
\big]= - 4 \pi \big(\overset{2}{T^{00}}\!-2\overset{2}g_{00} \overset{0}T{^{00}}\! + \overset{2}T{^{ii}}\big)~,\label{R004}
\end{eqnarray}
where $\overset{N}{\mathcal{R}}_{\mu\nu}$ denotes the term in $\mathcal{R}$ of order ${\bar{v}}^N/\bar{r}^2$~.~ $\partial_t\!\equiv\!\partial/\partial t$ and $\partial_i\!\equiv\!\partial/\partial x^i$ with $(t,\,x^i)$ being the four-dimensional coordinates. $\nabla^2$ is Laplace operator. \\
\indent Imposing the infinity boundary conditions that all fields should vanish, we can obtain the 1PN metric as follows 
\begin{eqnarray}
&& \overset{2}g_{00}= - 2\phi~,\label{g002}\\
&& \overset{2}g_{ij} = -2 \phi \delta_{ij}~,
\end{eqnarray}
\begin{eqnarray}
&& \overset{3}g_{0i} = \zeta_i~,\label{g0i3}\\
&& \overset{4}g_{00}= - 2\phi^2-2\psi~,\label{g004}
\end{eqnarray}
with
\begin{eqnarray}
&& \phi(t,\bm{x}) = - \int \frac{\stackrel{0~~}{T^{00}}\!\!(t,\bm{x}')}{|\bm{x}-\bm{x}'| } ~ d^3x'~,\label{phi0}\\
&& \zeta_i(t,\bm{x})  = - 4\int \frac{\stackrel{1~~}{T^{0i}}\!\!(t,\bm{x}')}{|  \bm{x}-\bm{x}'| } ~ d^3x'~,\label{zeta0}\\
&& \psi(t,\bm{x})  = -\int \Big[\frac{1}{4\pi}\partial_t^2 \phi(t,\bm{x}') + \stackrel{2~~}{T^{00}}\!\!(t,\bm{x}')+\stackrel{2~~}{T^{ii}}\!\!(t,\bm{x}')\Big]\frac{d^3x'}{|  \bm{x}-\bm{x}'| }~,\label{psi0}
\end{eqnarray}
where $\phi$ denotes Newtonian potential caused by the gravitational source's rest mass. $\zeta_i$ is the gravitational vector potential generated by the mass current, and is also called the gravitomagnetic field~\cite{KopeikinMashhoon2002,Sereno2002,WSperhake2004,Sereno2005,KopeikinFomalont2007}. $\psi$ denotes the second potential contributed by the gravitational source's kinetic and Newtonian potential energy as well as the second time derivative of Newtonian potential in the whole space.

Finally, the 1PN metric can be written as
\begin{eqnarray}
&& ds^2= -\big[1 + 2(\phi+\psi) + 2\phi^2\big]dt^2 +2\zeta_i dx^idt + \big(1-2 \phi\big)  dx_i dx^i~.\label{metric}
\end{eqnarray}

For a non-relativistic test particle, its 1PN dynamic equation can be written as~\cite{Weinberg1972}
\begin{eqnarray}\label{accelation1}
&& \frac{d \bm{v}}{dt} = - \nabla (\phi + 2 \phi^2 + \psi)  + 3 \bm{v}\, \partial_t \phi + 4 \bm{v}(\bm{v} \!\cdot\! \nabla) \phi - \bm{v}^2 \nabla \phi - \partial_t \bm{\zeta} + \bm{v} \!\times\! (\nabla \!\times\! \bm{\zeta})~,\label{dynamics}
\end{eqnarray}
where $\bm{v}$ denotes the velocity of the test particle.

All the above contents can be found in Weinberg's textbook~\cite{Weinberg1972}. The reason we repeat the derivation details of the post-Newtonian approximation theory is to show that Eqs.\,(\ref{phi0}) - (\ref{psi0}) hold in any harmonic coordinates, and we will make use of this argument in the later discussions.

So far we have only discussed the non-relativistic cases including the test particle. For a relativistic test particle, Eq.\,(\ref{metric}) can only provide its partial 1PN dynamic equation. For simplicity, here we only write down the complete 0.5PN dynamic equation of the relativistic test particle including photon as follow
\begin{eqnarray}\label{accelation2}
&& \frac{d \bm{v}}{dt} =  - (1+\bm{v}^2) \nabla \phi  + \bm{v} (3-\bm{v}^2) \partial_t\phi + 4\bm{v} (\bm{v} \!\cdot\! \nabla) \phi + \bm{v} \!\times\! (\nabla \!\times\! \bm{\zeta}) - \bm{v} (\bm{v} \!\cdot\! \nabla) (\bm{v}\!\cdot\!\bm{\zeta})~.\label{dynamics2}
\end{eqnarray}
where $\bm{v}$ denotes the velocity of the relativistic test particle and its magnitude is close to 1.

\section{The 1PN metric for a collapsing infinitely-thin spherical shell}\label{shell}

We consider a massive and pressureless infinitely-thin spherical shell, which has rest mass $M$ and radius $R$. The shell collapses under its self-gravity. The collapsing velocity is $un_i$ with $n_i$ being the radial unit vector, i.e., $\frac{dR}{dt}\!=\!u\!<\!0$, and $n_i\!=\!\frac{x_i}{r}$ with $r\!\equiv\!|\bm{x}|$. 
We calculate the metric of the self-gravitating and collapsing spherical shell to the 1PN order, so we only need the equation-of-motions for the shell at the Newtonian order, 
\begin{eqnarray}
&& \frac{du}{dt}=\frac{\Phi}{2R}~,\label{ND}
\end{eqnarray}
where $\Phi\!=\!\!-\frac{M}{R}$ is Newtonian potential induced by the other parts of the shell. Eq.\,(\ref{ND}) satisfies energy conservation in Newton theory -- the sum of Newtonian kinetic and potential energies of the shell per unit mass is constant, i.e.,  
\begin{eqnarray}
E_N \equiv \frac{1}{2}u^2-\frac{1}{2}\frac{M}{R} = constant ~.\nonumber
\end{eqnarray}

The energy-momentum tensor of the self-gravitating and collapsing thin spherical shell at the Newtonian order can be written as
\begin{eqnarray}
&&  \overset{0}T{^{00}}(t,\bm{x})= \frac{M}{4\pi R^2}\delta(r\!-\!R)~,\label{shellT000}\\
&& \overset{1}T{^{0i}}(t,\bm{x}) =  \frac{M}{4\pi R^2} \,u\, n_i \delta(r\!-\!R)~,\label{shellT0i1}\\
&&  \overset{2}T{^{00}}(t,\bm{x})= \frac{M}{4\pi R^2} \Big(\frac{1}{2}u^2+\Phi\Big)\delta(r\!-\!R)~,\label{shellT002}\\
&& \overset{2}T{^{ij}}(t,\bm{x}) =   \frac{M}{4\pi R^2} u^2 n_in_j\delta(r\!-\!R)~,\label{shellTij2}
\end{eqnarray}
where $\delta(s)$ is Dirac delta function of $s$. 

Substituting Eqs.~(\ref{shellT000}) and (\ref{shellT0i1}) into Eqs.\,(\ref{phi0}) and (\ref{zeta0}), it is easy to get the Newtonian potential and the gravitomagnetic potential of the shell as follows
\begin{equation}
\phi(t,\bm{x})= - \frac{M}{r}\theta(r\!-\!R)-\frac{M}{R}[1-\theta(r\!-\!R)]~,\label{phide}
\end{equation}
\begin{equation}
\zeta_i(t,\bm{x})= -\frac{4}{3}\frac{ M R \,u\, x_i}{r^3}\theta(r\!-\!R)-\frac{4}{3}\frac{ M \,u\, x_i}{R^2}[1-\theta(r\!-\!R)]~,\label{zetade}
\end{equation}
where $\theta(s)$ is the step function of $s$, which equals 1 for $s\!>\!0$ and 0 otherwise.

From Eq.\,(\ref{phide}) we can observe that the Newtonian potential inside the shell is dependent on time only and its gradient with respect to space is zero, so the internal gravitational field is zero at the Newtonian order, and this is in agreement with the prediction of Newton theory. However, beyond the Newtonian order, the dynamic equations of the test particles involve other potentials such as the second gravitational potential and the gravitomagnetic potential, as well as the time derivative of the Newtonian potential, as shown in Eqs.\,(\ref{dynamics}) and (\ref{dynamics2}). Later we will explicitly give the 1PN dynamic equation for the non-relativistic test particle and the 0.5PN dynamic equation for the relativistic one, both outside and inside the collapsing shell.


For the calculation of $\psi$, we need to get $\partial_t^2 \phi$ first. Taking the time derivative of Eq. (\ref{phide}), we have
\begin{eqnarray}
\partial_t \phi(t,\bm{x})=
\frac{M}{R^2}\dot{R}\big[1-\theta(r\!-\!R)\big]~,~\label{phi_shell_dot}
\end{eqnarray}
where ``dot" denotes the time derivative, and $\dot{R}$ is just $u$. Taking the time derivative of Eq. (\ref{phi_shell_dot}), we obtain
\begin{eqnarray}
&& \partial_t^2 \phi(t,\bm{x})  =  
 \Big(\!\! -2 \frac{M}{R^3}\dot{R}^2 + \frac{M}{R^2}\ddot{R}\Big)\big[1-\theta(r\!-\!R)\big]+\frac{M}{R^2}\dot{R}^2\delta(r\!-\!R) ~,\label{partialt20}
\end{eqnarray}
where we have made use of $\frac{d\theta(s)}{ds} \!=\! \delta(s)$.  \\

Then, substituting Eqs. (\ref{shellT002}), (\ref{shellTij2}) and (\ref{partialt20}) into Eq. (\ref{psi0}), after integration we can obtain the second gravitational potential of the shell as follow
\begin{eqnarray}
 \psi(t,\bm{x}) \!= \! -\frac{M}{r}\Big(\frac{11}{6}u^2\!\!+\!\frac{7}{6}\Phi\Big)\theta(r\!-\!R)\!-\!\frac{M}{R}\Big[\Big(\frac{3}{2}\!
+\!\frac{1}{3}\frac{r^2}{R^2}\Big)u^2\!
\!+\!\Big(\frac{5}{4}\!-\!\frac{1}{12}\frac{r^2}{R^2}\Big)\Phi
\Big]\!\big[1\!-\!\theta(r\!-\!R)\big].\label{pside}
\end{eqnarray}
where we have replaced $\dot{R}$ with $u$ and made use of $\ddot{R}\!=\!\frac{\Phi}{2R}$. 

Plugging Eqs.\,(\ref{phide}), (\ref{zetade}) and (\ref{pside}) into Eq.\,(\ref{metric}), we can write down the 1PN metric of the self-gravitating and collapsing thin spherical shell as follows
\begin{eqnarray}
&& \hskip -0.5cm ds^2= - \Big[1\!-\!\frac{2M}{r}\Big(1\!+\!\frac{11}{6}u^2\!-\!\frac{7}{6}\frac{M}{R}\Big)\!+\!\frac{2M^2}{r^2} \Big] dt^2  - \frac{8}{3}\frac{M R \,u \,\bm{x}\!\cdot\!d\bm{x}}{r^3}dt + \Big(1\!+\!\frac{2M}{r}\Big) d\bm{x}^2,\nn\\
&& \hskip 0cm \text{for } r>R~,\label{m1}
\label{externalmetric}
\end{eqnarray}
and 
\begin{eqnarray}
&& \hskip -0.5cm ds^2\!=  \!- \Big\{\!1\!-\!\frac{2M}{R}\Big[1\!+\!\frac{3}{2}u^2\!-\!\frac{9}{4}\frac{M}{R}
\!+\!\frac{1}{3}\frac{r^2}{R^2}\!\Big(\!u^2\!+\!\frac{1}{4}\frac{M}{R}\Big)\Big]\!\Big\} dt^2
\!-\!\frac{8}{3}\frac{M u \,\bm{x}\!\cdot\!d\bm{x}}{R^2}dt \!+\! \Big(1\!+\!\frac{2M}{R}\Big) d\bm{x}^2,\nn\\
&& \hskip 0cm \text{for } r\leq R~.\label{m2}
\label{internalmetric}
\end{eqnarray}
It can be seen that the metric is time dependent and continuous at the thin shell's position ($r\!=\!R$). 

In Appendix, we will not only prove that Eqs.\,(\ref{m1}) and (\ref{m2}) do satisfy Einstein field equations to the 1PN order, but also retrieve the thin spherical shell's energy-momentum tensor 
given in Eqs.\,(\ref{shellT000}) - (\ref{shellTij2}) from the time-dependent metric given by Eqs.\,(\ref{m1}) and (\ref{m2}). 

\section{The dynamic equations of the test particles}\label{dynamic equation}
Basing on 1PN metric, we can obtain the dynamic equations of the test particles in the spacetime of the collapsing spherical shell.

\subsection{Outside the shell}

Substituting Eqs.\,(\ref{phide}), (\ref{zetade}) and (\ref{pside}) into Eq.\,(\ref{accelation1}), we can write down the 1PN dynamic equation of the non-relativistic test particle outside the shell
\begin{eqnarray}\label{vnonoutside}
\frac{d\bm{v}}{dt}=-\frac{M\bm{x}}{r^3}\Big(1+E_N+\bm{v}^2-4\frac{M}{r}\Big)+\frac{4 M (\bm{v}\!\cdot\!\bm{x})\bm{v}}{r^3}~,
\end{eqnarray}
We have assumed that $|\bm{v}|$ has the same order of $|u|$. It follows this equation that the gravitational field outside the collapsing shell seems static to the 1PN order, since $E_N$ which is constant at the Newtonian order can be included into a re-defined mass.

For the case of the relativistic test particle with $|\bm{v}|\!\approx\!1$, the situation is different.
Substituting Eqs.\,(\ref{phide}), (\ref{zetade}) and (\ref{pside}) into Eq.~(\ref{accelation2}), we can obtain the complete 0.5PN dynamic equation of the relativistic test particle as follow
\begin{eqnarray}\label{lightoutside}
\frac{d\bm{v}}{dt}=-\frac{M\bm{x}}{r^3}(1+\bm{v}^2)+\frac{4 M (\bm{v}\!\cdot\!\bm{x})\bm{v}}{r^3}+\frac{4 M R\, u}{3r^3}\Big[\bm{v}^2\!-3\frac{(\bm{v}\!\cdot\!\bm{x})^2}{r^2}\Big]\bm{v}~.
\end{eqnarray}
In can be observed that the relativistic test particle does experience the time-varying effect induced by the radial motion, since the last term in Eq.~(\ref{lightoutside}) contains $R$ and $u$, and the time derivative of $R\,u$, which at the Newtonian order can be written as 
\begin{equation}\label{Ru}
\frac{d(R\,u)}{dt}
=E_N+\frac{1}{2}u^2~,
\end{equation}
does not equal zero in general. Notice that this term also appears in the dynamic equation of the non-relativistic test particle, but it is counted as a 2PN term there. 

\vskip -1cm 
\subsection{Inside the shell}

Similarly, from the gravitational potentials and gravitomagnetic field inside the shell,
we can write down the 1PN dynamic equation of the non-relativistic test particle
\begin{eqnarray}\label{vnoninside}
\frac{d\bm{v}}{dt}=-\frac{M\bm{x}}{R^3}\Big(2u^2+\frac{1}{2}\frac{M}{R}\Big)+\frac{3 M u }{R^2}\bm{v}~,
\end{eqnarray}
where we have assumed that $|\bm{v}|$ has the same order of $|\bm{u}|$.

For the relativistic test particle with $|\bm{v}|\!\approx\!1$,
we can write down the complete 0.5PN dynamic equation inside the shell as follow
\begin{eqnarray}\label{lightinside}
\frac{d\bm{v}}{dt}=\frac{3Mu}{R^2}\Big(1+\frac{1}{9}\bm{v}^2\Big)\bm{v}~.
\end{eqnarray}

In can be observed that the test particles' acceleration is zero at the Newtonian order. However, the 
test particles will experience the time-varying gravitational field beyond the Newtonian order. 
It is interesting to see that at the 0.5PN order, the relativistic test particle will keep moving straightly and decelerating since the value of $u$ is negative.

\section{Conclusion}\label{con}

Basing on the theory of the post-Newtonian approximation, we have calculated the metric of the self-gravitating and collapsing infinitely-thin spherical shell, and derived the test particles' equations-of-motion in the spherical shell's internal and external gravitational field. In order to check the correctness of the achieved metric, we have substituted it into Einstein field equation, and have retrieved the energy-momentum tensor of the thin spherical shell successfully. Our results demonstrate that the dynamic spherically-symmetric system can produce the time-dependent gravitational field.  


\appendix
\section{Proof for the time-dependent metric of the self-gravitating and collapsing thin spherical shell}




\begin{proof} 
From Eqs.\,(\ref{m1}) and (\ref{m2}), we can decompose the thin spherical shell's metric as follows
\begin{equation}
\overset{2}g_{00} = \left\{
\begin{array}{ll}
 2\frac{M}{r}~, ~~& \text{if } ~~r > R ,\\
\textcolor[rgb]{0.00,0.00,0.00}{ 2\frac{M}{R}~,} ~~& \textcolor[rgb]{0.00,0.00,0.00}{\text{if } ~~r \le R.}
\end{array} \right. \label{g002a}
\end{equation}
\begin{equation}
\overset{2}g_{ij} = \left\{
\begin{array}{ll}
 2\frac{M}{r}\delta_{ij}~, ~~& \text{if } ~~r > R~ ,\\
\textcolor[rgb]{0.00,0.00,0.00}{ 2\frac{M}{R}\delta_{ij}~,} ~~& \textcolor[rgb]{0.00,0.00,0.00}{\text{if } ~~r \le R~.}
\end{array} \right. \label{gij2a}
\end{equation}
\begin{equation}
\overset{3}g_{0i}  = \left\{
\begin{array}{ll}
 -\frac{4}{3}\frac{ M R \,u\, x_i}{r^3}~, ~~& \text{if } ~~r > R~ ,\\
 \textcolor[rgb]{0.00,0.00,0.00}{ -\frac{4}{3}\frac{ M \,u\, x_i}{R^2}~,} ~~& \textcolor[rgb]{0.00,0.00,0.00}{\text{if } ~~r \le R~.}
\end{array} \right. \label{g0i3a}
\end{equation}
\begin{equation}\label{g004a}
\overset{4}g_{00}  = \left\{
\begin{array}{ll}
2\frac{M}{r}\Big(\!\!-\!\frac{M}{r}+\frac{11}{6}u^2+\frac{7}{6}\Phi\Big)~, ~~& \text{if } ~~r > R~ ,\\
2\frac{M}{R}\Big[\!-\!\frac{M}{R}+\Big(\frac{3}{2}\!+\!\frac{1}{3}\frac{r^2}{R^2}\Big)u^2
+\Big(\frac{5}{4}\!-\!\frac{1}{12}\frac{r^2}{R^2}\Big)\Phi
\Big]~, ~~& \text{if } ~~r \le R~. 
\end{array} \right.
\end{equation}

Firstly, we show this metric satisfies the harmonic-coordinate conditions at the 1PN order, which can be written as~\cite{Weinberg1972}
\begin{eqnarray}
&& \frac{1}{2} \partial_t \overset{2}g_{00} - \partial_i \overset{3}g_{0i} + \frac{1}{2} \partial_t \overset{2}g_{ii}=0~,\label{h0a} \\
&& \frac{1}{2} \partial_i \overset{2}g_{00} + \partial_j \overset{3}g_{ij} - \frac{1}{2} \partial_i \overset{2}g_{jj}=0~.\label{hia}
\end{eqnarray}
From Eqs.\,(\ref{g002a}) - (\ref{g0i3a}), we have
\begin{eqnarray}
&& \partial_t \overset{2}g_{00} = -2\frac{Mu}{R^2}\big[1\!-\!\theta(r\!-\!R)\big]~,\label{h01a}\\
&& \partial_t \overset{2}g_{ii} = -6\frac{Mu}{R^2}\big[1\!-\!\theta(r\!-\!R)\big]~,\label{h02a}\\
&& \partial_i \overset{3}g_{0i} = -4\frac{Mu}{R^2}\big[1\!-\!\theta(r\!-\!R)\big]~,\label{h03a}\\
&& \partial_i \overset{2}g_{00} = -2 \frac{Mx_i}{r^3}\theta(r\!-\!R)~,\label{hi1a}\\
&& \partial_j \overset{2}g_{ij} = - 2\frac{Mx_i}{r^3}\theta(r\!-\!R)~,\\
&& \partial_i \overset{2}g_{jj} = -6 \frac{Mx_i}{r^3}\theta(r\!-\!R)~,\label{hi3a}
\end{eqnarray}
where we have replaced $\frac{dR}{dt}$ with $u$. With these equations, we can easily verify that Eqs.\,(\ref{h0a}) and (\ref{hia}) hold. 

Secondly, we verify that the 1PN metric satisfies Einstein field equations about $\stackrel{2}{\mathcal{R}}_{00}$,  $\stackrel{2}{\mathcal{R}}_{ij}$ and  $\stackrel{3}{\mathcal{R}}_{0i}$, which are given by Eqs.\,(\ref{R002}) - (\ref{R0i3}).

Plugging Eqs.\,(\ref{g002a}) - (\ref{g0i3a}) into Eqs.\,(\ref{R002}) - (\ref{R0i3}) respectively, we have 
\begin{eqnarray}
&&  \hskip -0.5cm \stackrel{2}{\mathcal{R}}_{00}=\frac{1}{2}\partial_i\partial_i\overset{2}g_{00}=
\frac{1}{2}\partial_i\Big[\!-\!2\frac{Mx_i}{r^3}\theta(r\!-\!R)\Big]\!\nn\\ 
&&  \hskip -0.5cm ~~~~~~ =-\frac{M}{r^2}\delta(r\!-\!R)= -4\pi \overset{0}T{^{00}}~,\label{R002a}%
\\
&&  \hskip -0.5cm \stackrel{2}{\mathcal{R}}_{ij}=\frac{1}{2}\partial_k\partial_k\overset{2}g_{ij} =\frac{1}{2}\partial_k\Big[\!-\!2\frac{Mx_k}{r^3}\theta(r\!-\!R)\delta_{ij}\Big]\nn\\ 
&& \hskip -0.5cm ~~~~~~=-\frac{M}{r^2}\delta(r\!-\!R)\delta_{ij}= -4\pi \delta_{ij}\overset{0}T{^{00}}~,\label{Rij2a}%
\\
&&  \hskip -0.5cm \stackrel{3}{\mathcal{R}}_{0i}=\frac{1}{2}\partial_j\partial_j\overset{3}g_{0i}=
\frac{1}{2}\partial_j\Big[
4MuR\Big(\frac{\delta_{ij}}{3R^3}\!-\!\frac{\delta_{ij}}{3r^3}\!+\!\frac{x_ix_j}{r^5}\Big)\theta(r\!-\!R)
\!-\!\frac{4Mu\,\delta_{ij}}{3R^2}\Big]\nn\\
&&  \hskip -0.5cm ~~~~~~= 2M u R \Big[\Big(\frac{n_i}{r^4}\!+\!3\frac{x_i}{r^5}\!+\!\frac{\delta_{ij}x_j}{r^5}\!-\!5\frac{x_i}{r^5}\Big)\theta(r\!-\!R)
\!+\!\Big(\frac{\delta_{ij}}{3R^3}\!-\!\frac{\delta_{ij}}{3r^3} \!+\! \frac{ x_ix_j}{r^5}\Big)n_j \delta(r\!-\!R) \Big]\nn\\
&&  \hskip -0.5cm ~~~~~~= \frac{2M u\, n_i}{ R^2}  \delta(r\!-\!R)= 8\pi \overset{1}T{^{0i}}~,\label{R0i3a}
\end{eqnarray}
where $\theta(s)$ is the step function of $s$, which equals 1 for $s\!>\!0$ and 0 otherwise. Here we have made use of $x_j\!=\!rn_j$, $\partial_j x_j\!=\!3$, $\partial_j x_i\!=\!\delta_{ij}$, $\partial_j r\!=\!n_j$ and $\frac{d}{ds}\theta(s)=\delta(s)$. Notice that $R$ and $u$ are the functions of time $t$ only. 

From Eqs.\,(\ref{R002a}) - (\ref{R0i3a}), we can recover $\overset{0}T{^{00}}$ and $\overset{1}{T}{^{0i}}$ for the collapsing thin spherical shell as follows
\begin{eqnarray}
&&  \overset{0}T{^{00}}= \frac{M}{4\pi R^2}\delta(r\!-\!R)~,\label{shellT000a}\\
&& \overset{1}T{^{0i}}=  \frac{M}{4\pi R^2} \,u\, n_i \delta(r\!-\!R)~,\label{shellT0i1a}
\end{eqnarray}
which are just Eqs.\,(\ref{shellT000}) and (\ref{shellT0i1}). 
In other words, the 1PN metric satisfies Eqs.\,(\ref{R002}) - (\ref{R0i3}). 

With Eqs.\,(\ref{shellT000a}) and (\ref{shellT0i1a}), we can directly write down $\overset{2}T{^{ij}}$ for the collapsing thin spherical shell 
\begin{eqnarray}
\overset{2}T{^{ij}} =   \frac{M}{4\pi R^2} u^2 n_in_j\delta(r\!-\!R)~,\label{shellTij2a}
\end{eqnarray}
which is just Eq.\,(\ref{shellTij2}).

Thirdly, we verify that the 1PN metric satisfies Einstein field equations about $\stackrel{4}{\mathcal{R}}_{00}$, which is given by Eq.\,(\ref{R004}). This equation   
is equivalent to~\cite{Weinberg1972} 
\begin{equation}
\nabla^{2}\psi=\partial_t^2 \phi +4 \pi \big(\overset{2}T{^{00}} +\overset{2}T{^{ii}}\big)~. \label{psisa}
\end{equation}
Notice that $\phi$ and $\psi$ are related to $\overset{2}g_{00}$ and $\overset{4}g_{00}$ via Eqs.\,(\ref{g002}) and (\ref{g004}). From $\overset{2}g_{00}$ and $\overset{4}g_{00}$ we have 
\begin{equation}
 \phi\!=\! - \frac{M}{r}\theta(r\!-\!R)-\frac{M}{R}\big[1-\theta(r\!-\!R)\big]~,\label{phidea}
\end{equation}
\begin{equation}
\hskip 0.5cm \psi \!= \! -\frac{M}{r}\Big(\frac{11}{6}u^2\!\!+\!\frac{7}{6}\Phi\Big)\theta(r\!-\!R)\!-\!\frac{M}{R}\Big[\Big(\frac{3}{2}\!
+\!\frac{1}{3}\frac{r^2}{R^2}\Big)u^2\!
\!+\!\Big(\frac{5}{4}\!-\!\frac{1}{12}\frac{r^2}{R^2}\Big)\Phi
\Big]\!\big[1\!-\!\theta(r\!-\!R)\big]~,\label{psidea}
\end{equation}
which are just Eqs.\,(\ref{phide}) and (\ref{pside}).

Taking the second time derivative of $\phi$ given by Eq.\,(\ref{phidea}), making use of $\frac{du}{dt}\!=\!\frac{\Phi}{2R}$ and replacing $\frac{dR}{dt}$ with $u$, we can confirm 
\begin{eqnarray}
 \partial_t^2 \phi =  -\frac{M}{R^3}
 \Big(2 u^2 \!-\! \frac{1}{2}\Phi\Big)\big[1\!-\!\theta(r\!-\!R)\big]+\frac{M}{R^2}u^2\delta(r\!-\!R) ~,\label{partialt2}
\end{eqnarray}
which is just Eq.\,(\ref{partialt20}). 

Applying Laplace operator on $\psi$ given by Eq.\,(\ref{psidea}), we can verify Eq.\,(\ref{psisa}) as follow
\begin{eqnarray}
&& \hskip -0.5cm \nabla^2 \psi = \partial_i\big(\partial_i \psi\big) \nn\\
&& = \partial_i \Big\{\Big(\frac{11}{6}u^2\!+\!\frac{7}{6}\Phi\Big)\frac{Mx_i}{r^3}\theta(r\!-\!R)\!
-\!\frac{M}{R^3}\Big(\frac{1}{3}u^2\!-\!\frac{1}{12}\Phi\Big)2x_i
\big[1\!-\!\theta(r\!-\!R)\big] \Big\}\nn\\
&&  = \Big(\frac{11}{6}u^2\!\!+\!\frac{7}{6}\Phi\Big)\frac{M}{R^2}\delta(r\!-\!R)\!-\!\frac{M}{R^3}\Big(2u^2\!\!-\!\frac{1}{2}\Phi\Big)
\big[1\!-\!\theta(r\!-\!R)\big]\!+\!\frac{M}{R^2}\Big(\frac{2}{3}u^2\!\!-\!\frac{1}{6}\Phi\Big)
\delta(r\!-\!R) \nn\\
&& =\underbrace{-\frac{M}{R^3}\Big(2u^2\!-\!\frac{1}{2}\Phi\Big)
\big[1\!-\!\theta(r\!-\!R)\big]\!+\!\frac{M}{R^2}u^2\delta(r\!-\!R)}_{\partial_t^2 \phi}\!+\!\underbrace{\frac{M}{R^2}\Big[\Big(\frac{1}{2}u^2\!+\!\Phi\Big)\!+\!u^2\Big]\delta(r\!-\!R)}_{ 4 \pi \big(\overset{2}T{^{00}}\,+\, \overset{2}T{^{ii}}\big)}.\label{nablapsi}
\end{eqnarray}
Notice that $\overset{2}T{^{00}}$ and $\overset{2}T{^{ij}}$ are given by Eqs.\,(\ref{shellT000a}) and (\ref{shellTij2a}) respectively, and $\partial_t^2 \phi$ is given by Eq.\,(\ref{partialt2}). Therefore, Eq.\,(\ref{psisa}) holds, implying the 1PN metric satisfies Eq.\,(\ref{R004}). 

Conversely, it can be retrieved from Eq.\,(\ref{nablapsi}) that 
\begin{eqnarray}
\overset{2}T{^{00}}+\overset{2}T{^{ii}}= \frac{M}{4\pi R^2}\Big(\frac{3}{2}u^2+\Phi\Big)\delta(r-R)~.\label{T00+Tii}
\end{eqnarray}
From Eq.\,(\ref{shellTij2a}), we have $\overset{2}T{^{ii}}\!=\!\frac{M}{4 \pi R^2}u^2\delta(r\!-\!R)$. Substituting it into Eq.\,(\ref{T00+Tii}), we can obtain $\overset{2}T{^{00}}$ for the collapsing spherical shell as 
\begin{eqnarray}
\overset{2}T{^{00}}=\frac{M}{4\pi R^2}\Big(\frac{1}{2}u^2\!+\!\Phi\Big)\delta(r\!-\!R)~, 
\end{eqnarray}
which is just Eq.\,(\ref{shellT002}). 

To sum up, we have not only proved that the time-dependent metric of the self-gravitating and collapsing thin spherical shell given by Eqs.\,(\ref{m1}) and (\ref{m2}) does satisfy Einstein field equations, but also retrieved all components of the shell's energy-momentum tensor, as given in Eqs.\,(\ref{shellT000}) - (\ref{shellTij2}). 

\end{proof}

\section*{Acknowledgements}
This work was supported in part by the National Natural Science Foundation of China (Grant No. 11973025) and the Program for New Century Excellent Talents in University (Grant No. NCET-10-0702).

\newpage
\bibliographystyle{spphys}       

\end{document}